\pdfoutput=1
\documentclass[useAMS,usenatbib,fleqn]{mn2e}
\usepackage{graphicx,subfig,lscape}

\newcommand{\kms}{km\,s$^{-1}$}
\newcommand{\vs}{$v \sin i$}
\newcommand{\teff}{$T_{\rm eff}$}
\newcommand{\lgg}{$\log\,{g}$}

\title[MDI of Ap Stars II: Next Generation MDI of $\alpha^2$ CVn]{Stokes $IQUV$ Magnetic Doppler Imaging of Ap stars \\ II: Next Generation Magnetic Doppler Imaging of $\alpha^2$ CVn\thanks{
Based on observations obtained at the Canada-France-Hawaii Telescope (CFHT) which is operated by the National Research Council of Canada, 
the Institut National des Sciences de l'Univers of the Centre National de la Recherche Scientifique of France,  and the University of Hawaii.  Also based on observations obtained at the Bernard Lyot Telescope (TBL, Pic du Midi, France) of the Midi-Pyr\'en\'ees Observatory,  which is operated by the Institut National des Sciences de l'Univers of the Centre National de la Recherche Scientifique of France. }
}

\author[J.Silvester, O. Kochukhov, G.A. Wade]
{J. Silvester$^{1,2}$, O. Kochukhov$^{3}$, G.A. Wade$^{2}$ \\
$^{1}$Department of Physics, Engineering Physics \& Astronomy, Queen's University, Kingston, Ontario, Canada, K7L 3N6\\
$^{2}$Department of Physics, Royal Military College of Canada, P.O. Box 17000, Station `Forces', Kingston, Ontario, Canada, K7K 7B4\\
$^{3}$Department of Astronomy and Space Physics, Uppsala University, 751 20, Uppsala, Sweden\\
}  

\begin{document}

\date{Accepted . Received }

\pagerange{\pageref{firstpage}--\pageref{lastpage}} \pubyear{2013}

\maketitle

\label{firstpage}

\begin{abstract}
We present updated magnetic field maps of the chemically peculiar B9p star $\alpha^2$ CVn created using a series of time resolved observations obtained using the high resolution spectropolarimeters ESPaDOnS and Narval. We compare these new magnetic field maps with the original magnetic Doppler imaging maps based on spectra recorded with the MuSiCoS spectropolarimeter and taken a decade earlier. These new maps are inferred from line profiles in all four Stokes parameters using the magnetic Doppler imaging code INVERS10. With the addition of new lines exhibiting Stokes $IQUV$ signatures we have a unique insight into how the derived magnetic surface structure may be affected by the atomic lines chosen for inversion.  We report new magnetic maps of  $\alpha^2$ CVn created using strong iron lines (directly comparable to the published MuSiCoS maps), weak iron lines and chromium lines, all of which yield a magnetic field structure roughly consistent with that obtained previously.

We then derive an updated magnetic structure map for $\alpha^2$ CVn based on the complete sample of Fe and Cr lines, which we believe to produce a more representative model of the magnetic topology of $\alpha^2$ CVn. In agreement with the previous mapping, this new updated magnetic map shows a dipolar-like field which has complex sub-structure which cannot be explained by a simple low order multipolar geometry.  Our new maps show that regardless of the atomic line or species choice, the reconstructed magnetic field is consistent with that published previously, suggesting that the reconstructed field is a realistic representation of the magnetic field of $\alpha^2$ CVn.  $\alpha^2$ CVn is the first Ap star for which multiple, high resolution magnetic maps have been derived, providing important observational evidence for the stability of both the large and small-scale magnetic field.
\end{abstract}


\section{Introduction}
The bright Ap star $\alpha^2$~CVn has been the subject of many observations over the past century, with the first period determination as early as Markov (1930), followed by Farnsworth (1932). There have been many studies of the magnetic and spectral variability of  $\alpha^2$~CVn (Babcock \& Burd (1952); Pyper (1969); Borra \& Landstreet (1977)). It was not until Kochukhov et al. (2002) employed a new magnetic Doppler imaging technique (MDI) (described by Piskunov \& Kochukhov (2002) and Kochukhov \& Piskunov (2002)) that the first high resolution maps of the surface vector magnetic field using Stokes $IV$ observations were made for $\alpha^2$~CVn. These maps were later refined by using linear polarisation profiles (Stokes $Q$ and $U$) in combination with Stokes $IV$ (Kochukhov \& Wade 2010) acquired with the MuSiCoS spectropolarimeter. These maps (along with those of 53 Cam, Kochukhov et al. 2004) were distinguished from earlier models in that they were computed directly from the observed polarised line profiles, making no {\em a priori} assumptions regarding the large-scale or small-scale topology of the field. The MDI surface magnetic field maps of both stars revealed that their magnetic topologies depart significantly from low-order multipoles.

These original maps were limited by the quality of the observational data. With MuSiCoS being a relatively inefficient medium resolution spectropolarimeter, only a very small number of lines exhibited Stokes $QU$ profiles of sufficient quality for modelling. With the new observations of Ap stars in all four Stokes $IQUV$ parameters obtained using the new ESPaDOnS and Narval instruments as described by Silvester et al. (2012), it is now possible to not only study more spectral lines, these new data also allow the study of $\alpha^2$~CVn at a resolution not previously possible and allow a more detailed probe of subtle spectral features which have been unresolved or buried in the noise in the MuSiCoS observations. 

The mapping performed by Kochukhov \& Wade (2010) (herein referred to as K\&W) used the inversion code INVERS10, which is also used in this study. INVERS10 employs a single mean metallicity model atmosphere when performing the inversions. It was suggested by Stift et al. (2012) that using a single mean model atmosphere, which did not account for horizontal (local) atmospheric variations would lead to the derivation of incorrect abundance distributions and incorrect magnetic field geometries. To address these concerns, Kochukhov et al. (2012) compared magnetic field maps of $\alpha^2$~CVn reconstructed with INVERS10 with those reconstructed using a version of the INVERS code which incorporated horizontal variation of the model atmosphere. Kochukhov et al. (2012) found no significant differences between the mapping results from the two codes, confirming the suitability of INVERS10 for mapping both the magnetic field and chemical abundance features in Ap stars such as $\alpha^2$~CVn. This work aims to further explore the results of Kochukhov at al. (2012), by investigating whether the same {\it unique} magnetic field topology can be obtained from various sets of suitable Stokes $IQUV$ lines, taken with the higher spectral resolution data from ESPaDOnS and Narval. 

\begin{table}
\caption{Fundamental parameters used/derived for  the $\alpha^2$ CVn mapping.  {\bf References}: (1) Kochukhov et al. (2002), (2) Farnsworth (1932), (3) Kochukhov and Wade (2010)}
\begin{tabular}{ccc}
\hline
\hline  
Parameter & Value & Reference \\
\hline
\teff &  $11600 \pm 500$ K  & (1) \\
\lgg & $3.9 \pm 0.1$ & (1) \\ 
$P_{rot}$ & 5.46939 days & (2) \\
\vs & $18.0 \pm  0.5$ \kms& \\
$i$ & $120^{\circ} \pm 5$ & (3) \\
$\Theta$ & $115^{\circ} \pm 5$ & (3) \\
\hline
\label{parameter-table}
\end{tabular}
\end{table}

The remainder of the paper is organised as follows: section 2 briefly describes the observations, section 3 discusses the procedure for selecting lines suitable for inversion. In section 4 we discuss the various magnetic maps of $\alpha^2$~CVn, with the results and implications of these maps. Finally we summarise our findings and the implications in the conclusion. 

\section{Spectropolarimetric observations}
Observations of $\alpha^2$ CVn were obtained between 2006 and 2010 with both ESPaDOnS and Narval spectropolarimeters during the observing campaign as described by Silvester et al. (2012). In total 28 Stokes $IQUV$ observations of  $\alpha^2$ CVn  were obtained. The reduction of observations was carried out at the observatories using the dedicated software package Libre-ESpRIT which yields both the $I$ spectrum and the $V$ circular polarisation spectrum and/or $QU$ polarisation spectra of each star observed. In this work each reduced spectrum is normalised order-by-order using an IDL code specifically optimised to fit the continuum of these stars. The full details of the observations and reduction are reported by Silvester et al. (2012), along with the log of observations for $\alpha^2$ CVn. Importantly Silvester et al. (2012) showed that the resulting longitudinal magnetic field and net linear polarisation measurements obtained with ESPaDOnS and Narval were consistent with those measured from with MuSiCoS spectra obtained by Wade et al. (2000), making them suitable for direct comparison.

\begin{table}
\caption{Atomic lines used for  the $\alpha^2$ CVn mapping. The $\log gf$ values are those as provided by the Vienna Atomic Line Database (VALD, Kupka et al. 1999)}
\begin{tabular}{ccc}
\hline
\hline  
Ion & Wavelength & $\log gf$ \\
&(\AA )  & \\
\hline
Cr II &  4824.127  &  -1.085 \\
         & 5246.768 & -2.560 \\
         & 5279.876 & -2.112 \\
         & 5280.054 & -2.316 \\
\hline
Fe II & 4273.326 &  -3.303 \\
	& 4520.224 & - 2.617 \\
          & 4666.758 &  -3.368 \\
          & 4923.927 &  -1.320 \\
          &  5018.440 &  -1.220 \\
          & 5019.462 & -2.784 \\        
\hline
\label{line-list}
\end{tabular}
\end{table}

\begin{figure*}
\begin{center}
  \includegraphics[width=0.85\textwidth]{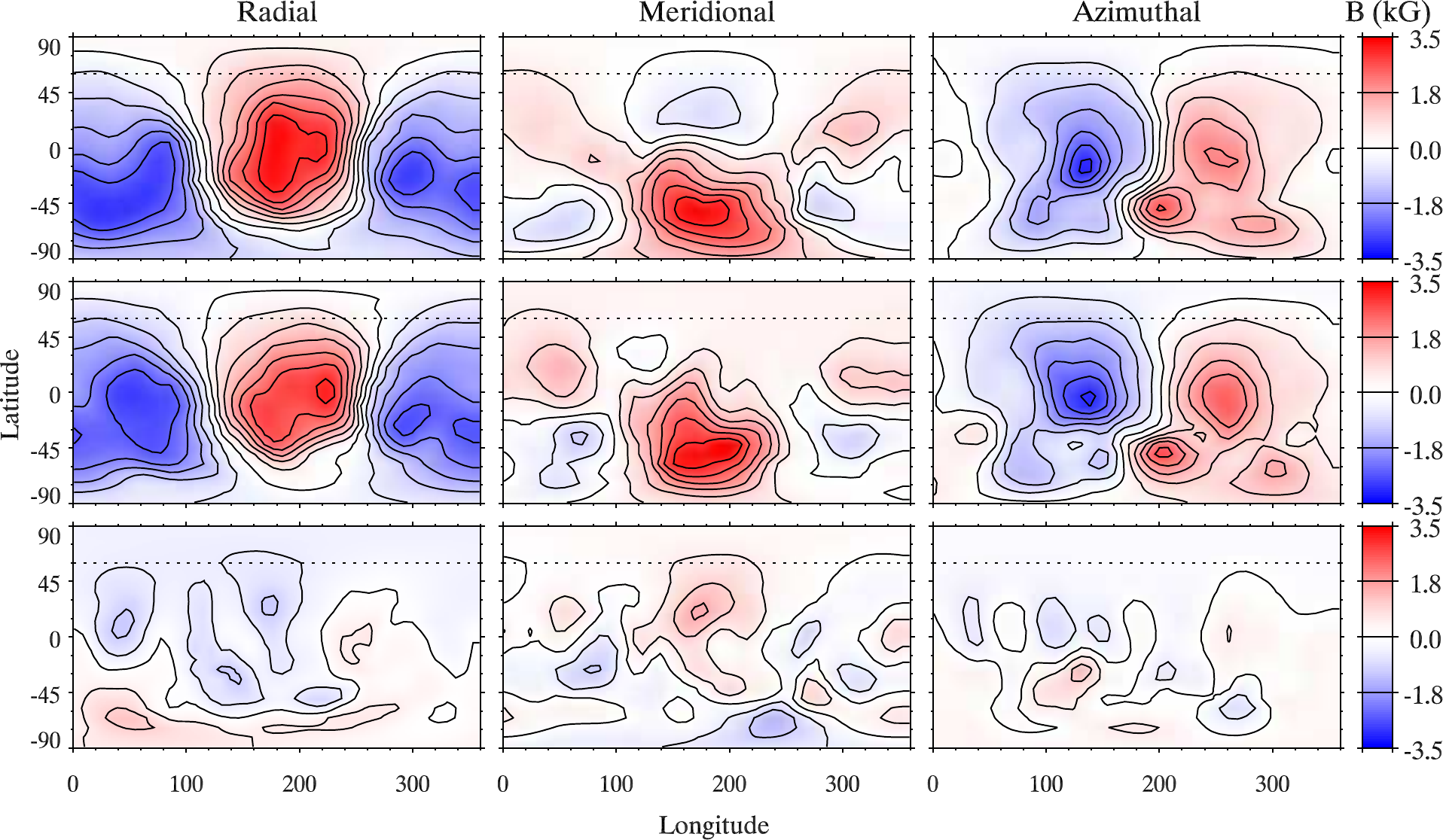}
   \caption{Comparison of magnetic field radial, meridional and azimuthal components derived from MDI maps computed using strong Fe lines, presented in rectangular projection. Upper row: results from the MuSiCoS dataset of K\&W. Middle row: results from the ESPaDOnS/Narval dataset of Silvester et al. (2012). Lower row: difference maps corresponding to the middle row minus the upper row.  Dashed line indicates the highest possible visible latitude based on the adopted inclination angle $i =120^{\circ}$.  A contour stepping of 0.5 kG has been used for a range of [-3.5,+3.5] kG. }
\label{field-maps-rec}
\end{center}
\end{figure*}

\begin{figure*}
\begin{center}
  \includegraphics[width=0.85\textwidth]{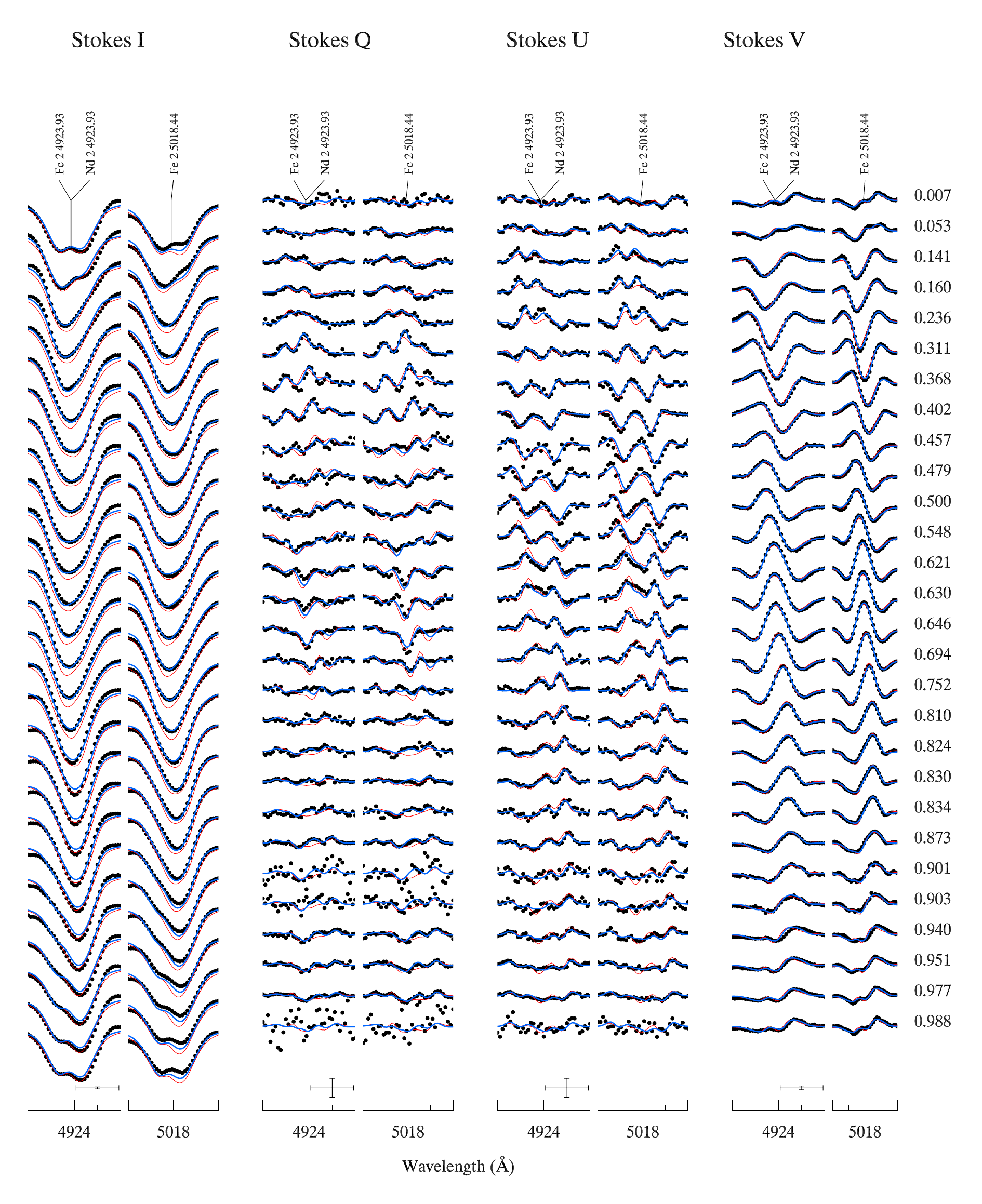}
   \caption{Stokes $IQUV$ spectra comparison for strong iron lines;  observed (dots), synthetic from ESPaDOnS and Narval data (solid curve, blue) and from the MuSiCoS map (solid curve, red) for  $\alpha^2$ CVn.}
\label{Fe-fit-IQUV-Strong}
\end{center}
\end{figure*}

\section{Inversion}
The MDI mapping is performed using the INVERS10 code (as described by Piskunov \& Kochukhov (2002), Kochukhov \& Piskunov (2002) and K\&W). INVERS10 constructs model line profiles based on an assumed initial spherical surface distribution of free parameters (element abundance and magnetic field geometry) and then iteratively adjusts the parameters until the computed line profiles are in agreement with the observations. Mathematically MDI is a least-squares minimization problem where the polarised radiative transfer depends on the magnetic field and abundance distributions, with a regularization parameter which facilitates convergence and forces the code to choose the simplest of a possible multitude of solutions providing a good fit to the observations. 

The code is written in FORTRAN, with the graphical output and file processing performed under IDL. This code is fully parallelized and was run on a 8 CPU Mac Pro. The time required for the code to converge to a solution for a set of 4 Stokes parameter data (containing 28 phases), with fitting to one or two spectral lines, is of the order of a few hours. During the reconstruction of the magnetic field using INVERS10 we adopted the same value of temperature and effective gravity as used by K\&W.  We did however adopt a \vs\  of 18.0 \kms, which differs slightly from K\&W, who adopted a value of  18.4 \kms.  We found that during our mapping runs a value of 18.0 \kms\  gave slightly better agreement between the observation and model fit. This adopted value still agrees with K\&W within the error bars.  Table \ref{parameter-table} summarises the key stellar parameters used in this study.  

Another important parameter in the reconstruction of the magnetic field topology is the choice of the regularisation value within INVERS10. As is described by K\&W the Tikhonov regularisation function assists the code in converging to a solution by providing a limit on how smooth or patchy the resulting map can be.  Providing a reasonable amount of regularisation is a balance of making sure that the map is not so smooth as to ignore the fitting of discrete features in the observed spectra, whilst making sure the map is not allowed to become so patchy that the code has started to fit to the noise.  For the purpose of this work, a value of regularisation was chosen which gave the lowest total discrepancy, whilst still reproducing the Stokes profiles without fitting to a significant amount of noise.  When performing inversions to obtain a magnetic field map from a given element, it is also required that we constrain the abundance map for that element.  While abundance maps for both iron and chromium are derived in our analysis, we decided that the discussion of the interplay between the magnetic field and chemical atmospheric structure is beyond the scope of this paper.  Therefore, in this paper we will concentrate on the investigation of the robustness of the magnetic field diagnosed using MDI. A future paper will describe the abundance maps and their relationships to the magnetic field.

It is important to note that the MuSiCoS observations employed by K\&W were based on 15 complete (full Stokes $IQUV$) and 5 partial (one or more Stokes parameters missing) phases of observation,  whilst the new dataset is based on 28 complete phases.  The phase coverage of the MuSiCoS time series is generally good, but because of the lower number of complete phases there are some gaps (e.g. between phase 0.499 and 0.582 and between 0.594 and 0.706). With the new data there are fewer gaps although there is one between phase 0.053 and 0.141. Overall the phase sampling is very similar between the two data sets, making it reasonable to perform a comparison between magnetic field maps reconstructed using the same atomic lines between the two data sets. 
 
For magnetic mapping only lines which show clear signatures in Stokes $Q$ and $U$ were selected. The basis of line selection was to start with the lines used by Kochukhov et al. (2002) and then to expand the list using a visual inspection to pick lines in which signatures were clearly present compared to the noise. Lines which did not show variability or were heavily blended with other lines,  in addition to lines which suffered from significant non-LTE effects were avoided. As described in section 3 of Silvester et al. (2012), the ESPaDOnS and Narval spectra offered a much larger resolution than the MuSiCoS spectra  with $R = \lambda / \Delta \lambda \simeq 65000$), and wavelength coverage from 3690-10480 \AA\  (with small gaps at 9224 to 9234 \AA\, 9608 to 9636 \AA\,  and 10026 to 10074 \AA). Even with this improved data specification, the number of suitable lines based on the aforementioned criteria was only somewhat larger than the MuSiCoS set, although the new data were still of superior quality (The median signal-to-noise ratio of the reduced observations is 1000 per 1.8~\kms\ pixel vs 580 per 2.6~\kms\ pixel for the MuSiCoS data set).

\section{Results - Magnetic maps}

The magnetic maps from the new data set are based on various line sets; first for direct comparison with the MuSiCoS data the magnetic field was mapped using only the strong iron lines Fe~{\sc ii}\ $\lambda$  4923 and $\lambda$ 5018 to determine if by mapping the same lines using this newer data consistent results were obtained. This will be described in section 4.1.  We then identified additional lines which could be used to create magnetic Stokes $IQUV$ maps.  These additional lines allow us to determine if the reconstructed magnetic field topology depends on factors such as the line choice, the atomic species and line strength. To investigate this we created a Stokes $IQUV$ magnetic map based on weak iron lines (lines not used in previous mappings of  $\alpha^2$ CVn), then for all the iron lines combined and finally for chromium lines. These experiments are described in sections 4.2, 4.3 and 4.4 respectively. The ultimate culmination of this search for additional lines was a magnetic field map computed using both iron and chromium lines of which the results are described in section 4.6.  
In section 4.5 we compare a sample of observed profiles with synthetic profiles obtained from pure dipolar and dipolar + quadrupolar geometries. A list of the lines used is presented in Table \ref{line-list}, along with the the $\log gf$ used which was taken from the Vienna Atomic Line Database (VALD, Kupka et al. 1999).

\subsection{Strong iron line maps}
The first step in mapping with the new data was to produce a magnetic field map obtained from the strong iron lines of Fe~{\sc ii}\ $\lambda$ 4923 and 5018 as used in the original MuSiCoS maps of K\&W. This was to test if the new data produced a field topology consistent with that found by K\&W. To create the new map we used the same atomic data and model atmosphere used by K\&W. The only differences in the input file were the choice of \vs, as described in section 3, and the value of regularisation, which was adopted independently of the value used by K\&W. Because these new maps are created using a new data set, it is natural to have to independently adopt a value of regularisation. It should be noted that K\&W also used Cr~{\sc ii}\ $\lambda$ 4824 in the mapping of $\alpha^2$ CVn which we have not included in this reconstruction. This line will be included in later reconstructions (sections 4.4 and 4.6).

 A direct comparison of the radial, meridional and azimuthal fields derived from the two the data sets is shown in Fig. \ref{field-maps-rec} in a rectangular representation. Good agreement can be seen from the two data sets for all three field components, with the difference plot showing very little in the way of structure. The most significant discrepancies are in the magnitude of the meridional field, with differences on the order of 500 G. We consider this to be a result of the new data giving us a more precise measure of the magnetic field structure in these regions.  Given that the code uses Tikhonov regularization when performing magnetic inversions, it is reasonable to expect that the map derived from lower resolution data will have smaller peak magnetic field amplitudes to avoid fitting to the noise. It has been shown in studies such as Brown et al. (1991), Kochukhov \& Piskunov (2002) and Ros\'{e}n \& Kochukhov (2012) that the meridional field is the most difficult to constrain by inversions. Therefore it is not surprising to find the maximum difference appear in this particular field component. It is interesting to note that the iron abundance map that we derive from the strong iron lines is in very good agreement with that found by K\&W. This abundance map will be described in the next paper. 

The agreement between the synthetic profiles corresponding to the map and the observations is illustrated in Fig. \ref{Fe-fit-IQUV-Strong}. The agreement between the observed and synthetic profiles is good, with most of the features reproduced in all Stokes parameters without fitting to the noise.  Within Fig. \ref{Fe-fit-IQUV-Strong} we have also included the synthetic profiles from the MuSiCoS map (shown in red). Whilst the general shape is comparable to the new data fit, it is clear that the MuSiCoS fit does not fully account for the detailed structure of the new Stokes $Q$ and $U$ profiles. These discrepancies may reflect the differences observed in the maps of the meridional field component. It should be noted that whilst the MuSiCoS derived model does provide a good fit to the observed MuSiCoS Stokes $Q$ and $U$ profiles,  by reducing the resolution of the new observations to the resolution of MuSiCoS and performing a direct comparison, we find that new observations do indeed lead to a smaller discrepancy between the observations and model in Stokes $Q$ and $U$ compared with the fit obtained in K\&W, this suggests the new data contains new information.

Even with these differences,  there is good agreement between the two data sets obtained with a separation of over a decade. Thus this new map (herein referred to as the strong line map) is suitable as a basis for comparison to see how the magnetic map may differ depending on the choice of atomic lines used in the mapping by comparison with new magnetic maps created using different line sets.

\begin{figure*}
\begin{center}
\includegraphics[width=0.85\textwidth]{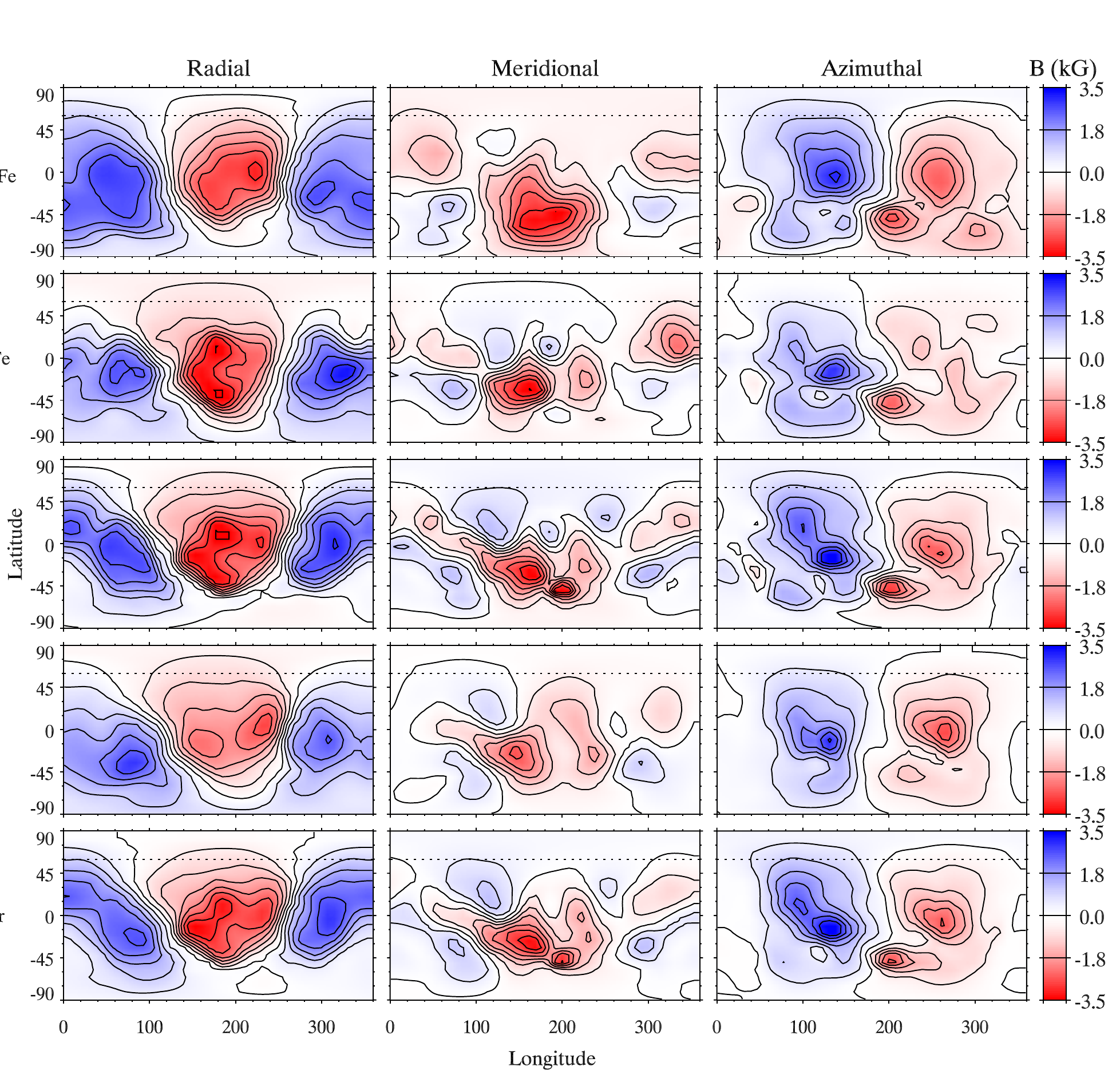}
   \caption{Rectangular maps of three magnetic field vector components for:  (from top to bottom) strong Fe lines, weak Fe lines, combined Fe line set, Cr lines and all lines. For information on figure details (dashed line, contours) refer to Fig. 2.}
\label{Allrecmaps}
\end{center}
\end{figure*}

\subsection{Weak iron line maps}
The next step was to investigate the effect on the reconstructed magnetic field topology by only using weak iron lines in the inversion.  This is the first time  $\alpha^2$ CVn has been mapped using weak intensity iron lines, because the previous mapping using MuSiCoS spectra did not present any weak iron lines with clear linear polarisation signatures (Kochukhov et al. 2002, K\&W).  Whilst weak iron lines have reduced Stokes $VQU$ signature amplitudes compared to the strong iron lines,  the weak lines are more sensitive to horizontal abundance variations and therefore maybe more suitable for reconstruction of chemical abundance maps. The question is can such weak lines still be used for accurate reconstruction of the magnetic topology. 

The weak iron lines used were Fe~{\sc ii}\ $\lambda\lambda\lambda$ 4273,  4520 and 4666. These lines were selected on the basis that they exhibited Stokes $QU$ profiles which had amplitudes clearly above the noise,  that the intensity profile showed significant variability (which is more important for abundance mapping) and that these lines had predicted line depths smaller than the strong iron line set. Whilst the spectrum of $\alpha^2$ CVn contains many iron lines,  most of the lines were strongly blended with other elements or did not show clear Stokes $QU$ profiles, which is why we are limited to three weak iron lines.  The  $\log gf$ values adopted in the inversion calculations are indicated in Table 2. 

A comparison between the resulting magnetic topology map (herein referred to as the weak line map) and the strong iron line map is shown (with other maps to be described in the following sections) in rectangular form in Fig. 3. By comparing the strong and weak iron line maps (first and second row respectively) it can be said that overall, the geometry of the derived field is quite similar, particularly in the radial field component, but there are some differences in the azimuthal structure and in particular there is a difference at large negative latitudes in the meridional field.  This could well be a result of the difference in sensitivity to horizontal structure between the two maps. Strong saturated lines change due to abundance spots over the stellar surface to a lesser extent than weak lines, with strong lines forming over a wider range of abundances than the weaker lines. This means the strong iron line map potentially represents a smoothed version of the field topology. In additional, as has already been mentioned the meridional field is the least constrained of the three field components, so it is not surprising that this component differs between the two maps. 

It is important to note that the weak line Stokes $Q$ and $U$ signatures are often of very different shape than the strong line signatures, probably due to their different sensitivities to the horizontal abundance structure, but also due to the different relative strengths of the $\pi$ and $\sigma$ components as a result of different levels of saturation. If we compute profiles of the weak lines using the strong line magnetic field map, the agreement is quite poor.  The same is also found if the reverse comparison is performed. In addition we also checked that these differences were not a consequence of differences in the abundance distribution maps. 

This shows that relatively small (but still non-negligible) differences in the magnetic maps can produce quite important differences in the resultant Stokes profiles. This in itself suggests that combining lines of different characteristics will probably result in a map which is more representative of the average field distribution. Also, the validity of the derived field map is tested through this process, since a single field model should be capable of reproducing the profiles and variations of all lines.

\subsection{Combined iron line maps}
The next logical step was to see the effect on the inferred magnetic field by combining the two iron line sets (herein referred to as the combined iron line map).  We combined the strong iron lines of Fe~{\sc ii}\ $\lambda$ 4923 and 5018, in addition to weaker iron lines of Fe~{\sc ii}\ $\lambda$ 4273, 4520 and 4666. Considering the non negligible differences between the two maps computed using the strong and weak lines, we were curious to investigate if a single magnetic field model was able to reproduce all lines simultaneously. In this map the two sets of lines are combined without applying any relative weights. The resulting magnetic map is shown in rectangular form compared to the strong iron line magnetic map in Fig. 3. By comparing the two maps (row one and row 3), once again we see that the overall field structure is consistent with the strong iron line magnetic field map, with the meridional field showing the largest differences at negative latitude. The differences are a result of differences in sensitivity to horizontal structure, as described in the previous section. 

For an indication of how well the inversion code has simultaneously fit both the strong and weak iron lines together,  Fig. \ref{CrFe-fit} illustrates the fit between observed profiles and model for the final map which includes the combined iron lines and chromium.  The fit of the iron lines in this figure is representative of the fit obtained for the combined iron line map, with good agreement in all Stokes parameters.

\subsection{Chromium line map}
Next we wanted to investigate the results when using chromium lines to produce Stokes $IQUV$ magnetic maps, again to see if the choice of line and in this case using a different element, modifies the resulting map when compared to the original strong iron line map.  Using the same criteria as used for the weak iron lines, the chromium lines of Cr~{\sc ii}\ $\lambda\lambda\lambda~4824$,  5246 and 5280 were chosen.  These lines have predicted depths much lower than the weak and strong iron lines.  The  $\log gf$ adopted in the inversion calculations are indicated in Table 2. The magnetic map calculated from the chromium lines is compared to the strong iron line magnetic map in Fig. 3.   The general structure between the two maps is in agreement.  There are some differences in particular the meridional component at negative latitude structure.  This is similar to what is seen in the weak iron line map (section 4.2).

\begin{figure*}
\begin{center}
  \includegraphics[width=0.85\textwidth]{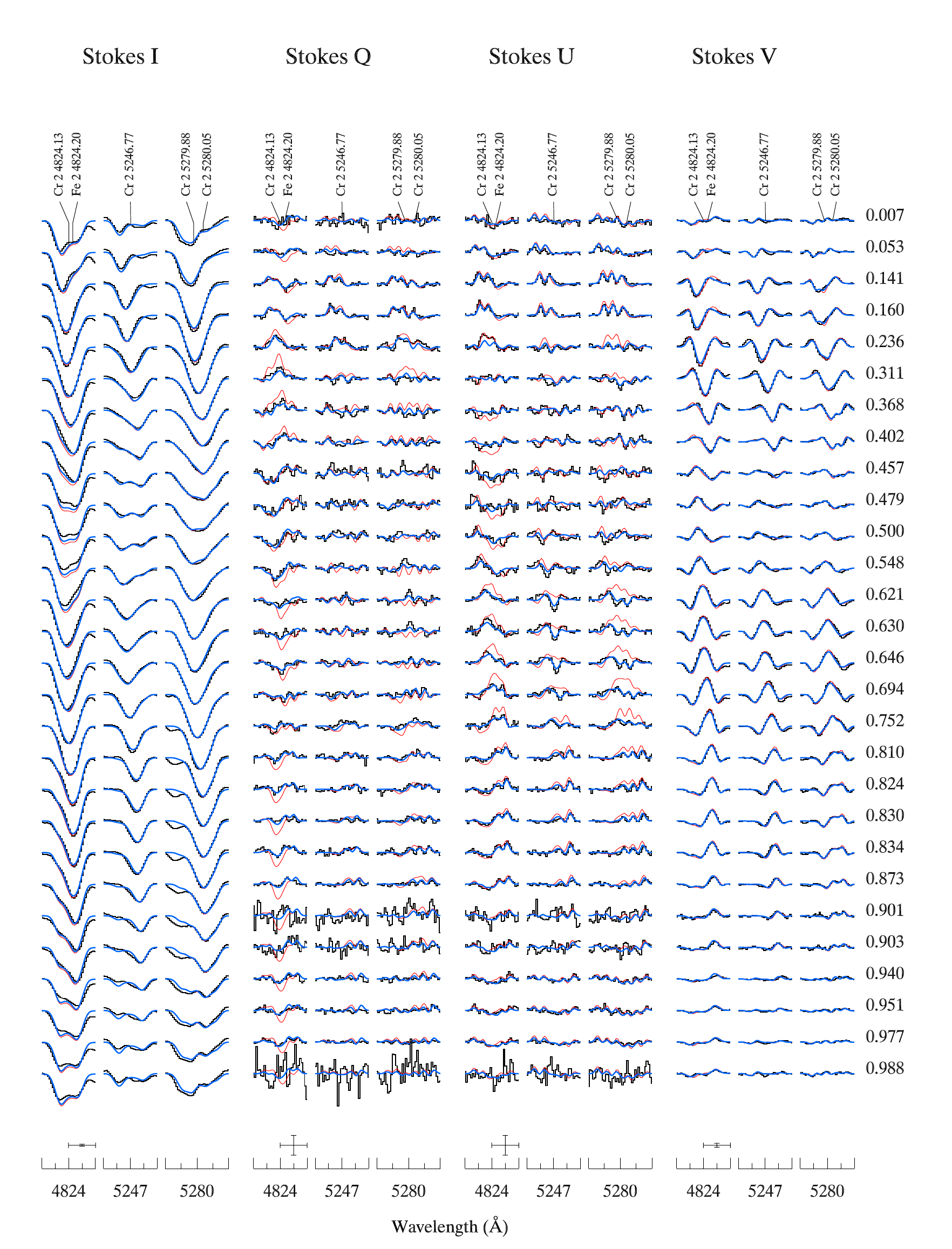}
   \caption{Comparison between observed chromium lines (dots) and synthetic profiles based on the chromium lines (solid curve, blue). Over plotted is the strong iron line magnetic field map with the chromium abundance synthetic profiles shown for comparison (solid curve, red).}
\label{Cr-fit-Fe-field}
\end{center}
\end{figure*}

The source of this difference can be seen in Fig. \ref{Cr-fit-Fe-field} which shows the observed chromium Stokes $IQUV$ profiles compared with synthetic profiles obtained from the chromium lines and the equivalent profiles obtained from the strong iron line magnetic map. It is important to note that these synthetic iron line profiles were computed using the chromium abundance map as found from the aforementioned chromium inversion.  This is required to eliminate any line profile differences caused by the differing iron and chromium chemical abundance structures.  As was mentioned in section 3, the abundance maps will be presented in a future paper. 

Within Fig. \ref{Cr-fit-Fe-field} the observed and computed Stokes $IV$ profiles are in good agreement, however there are differences in Stokes $QU$ with most profile features not being reproduced correctly by the model. This difference would lead to a difference in structure between the two maps. Again this difference we believe is the result of the different horizontal structure sensitives between the strong iron lines and the chromium lines. We believe that we are seeing a more representative model of the field by choosing weak iron lines or chromium lines, in which we are seeing the magnetic field as it is without the smoothing effects of the strong iron lines.

\subsection{Comparison to multipolar geometries}
As has been described in previous mapping of $\alpha^2$ CVn (K\&W, Kochukhov et al. 2012), the magnetic field topology shows complex substructure which could not be described by a low-order multipolar geometry.  With the new data it is important to examine whether or not this is still the case and evaluat if the observed profiles cannot be satisfactorily fitted with a more simple field topology.  To investigate this, the profiles for the combined iron line set were compared to model profiles one would obtain for an optimal dipolar and dipolar plus quadrupolar geometry.  To accomplish this comparison we have fitted four Stokes parameter observations with a modified version of INVERS10 in which a direct model description of the three field components was substituted with a multipolar parameterization similar to the one described by Donati et al. (2006). In this comparison the chemical abundance distribution has been allowed to vary as is the case with the other inversions performed. Further details about our implementation of MDI with multipolar expansion are provided by Kochukhov et al. (2013). In the present study we performed inversions using only poloidal field components and limiting multipolar expansion to angular degrees $\ell=1$ and $\ell=2$ for the dipolar and dipolar+quadrupolar models, respectively. The latter field parameterization is mathematically equivalent to the non-axisymetric dipolar plus quadrupolar model geometry employed by Bagnulo et al. (2002) in their statistical study of Ap star magnetic fields.

\begin{figure*}
\begin{center}
 \includegraphics[width=0.65\textwidth, angle=90]{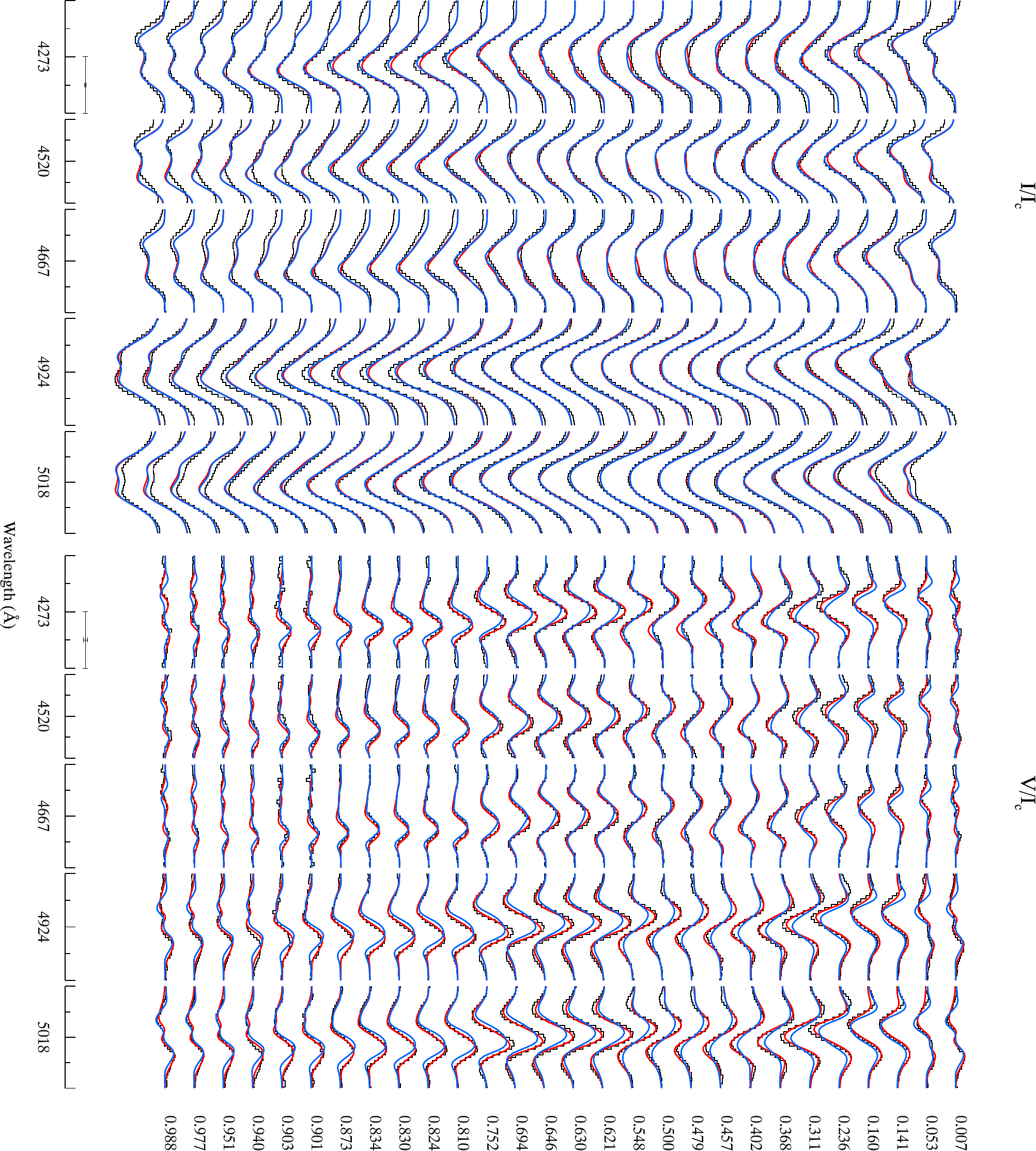}
    \includegraphics[width=0.65\textwidth, angle=90]{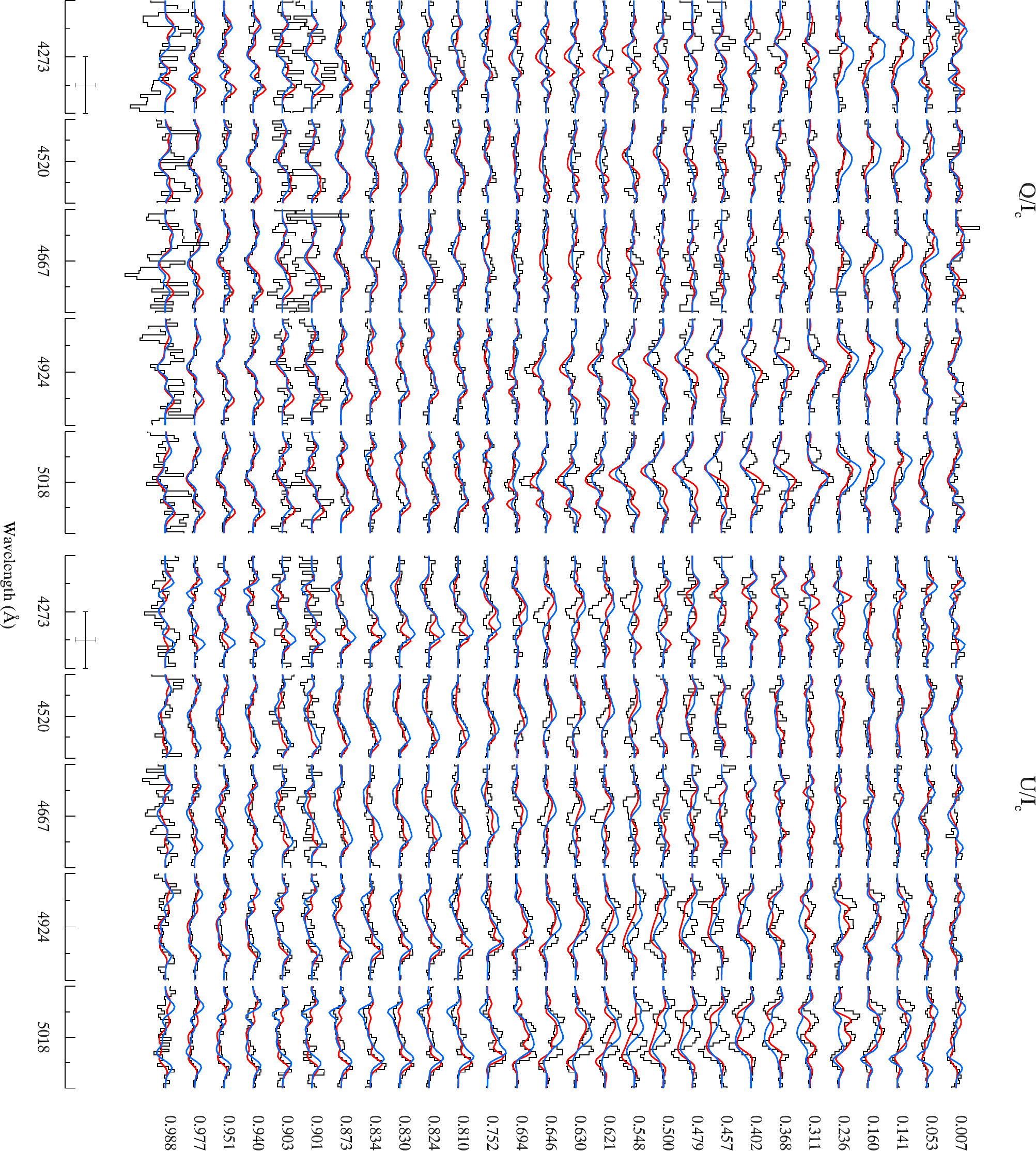}
   \caption{Comparison between observed (dots) for the weak and strong iron lines used in magnetic mapping and synthetic spectra for a dipolar geometry (solid curve, blue) and for a dipole + quadrupole geometry (solid curve, red). Upper frame: Stokes $I$ and $V$ profiles. Lower frame: Stokes $Q$ and $U$ profiles.}
\label{othergeofit}
\end{center}
\end{figure*}

As illustrated in Fig. \ref{othergeofit}, it can be seen that neither model provides good agreement with the profiles. The dipole model fails to fully reproduce Stokes $V$ and does not reproduce Stokes $Q$ and $U$. The dipolar+quadrupolar model does a better job of Stokes $V$, in the case of both Stokes $Q$ and $U$ it is clear the structure of the model generally fails to reproduce the high-contrast wavelength variation of the observed profile and it fails most dramatically at those phases where we infer the most significant complex structure to be visible. We can therefore conclude that a simple field topology cannot describe the field structure of $\alpha^2$ CVn, a result in agreement with the findings of K\&W.

\begin{figure*}
\begin{center}
    \includegraphics[width=0.85\textwidth]{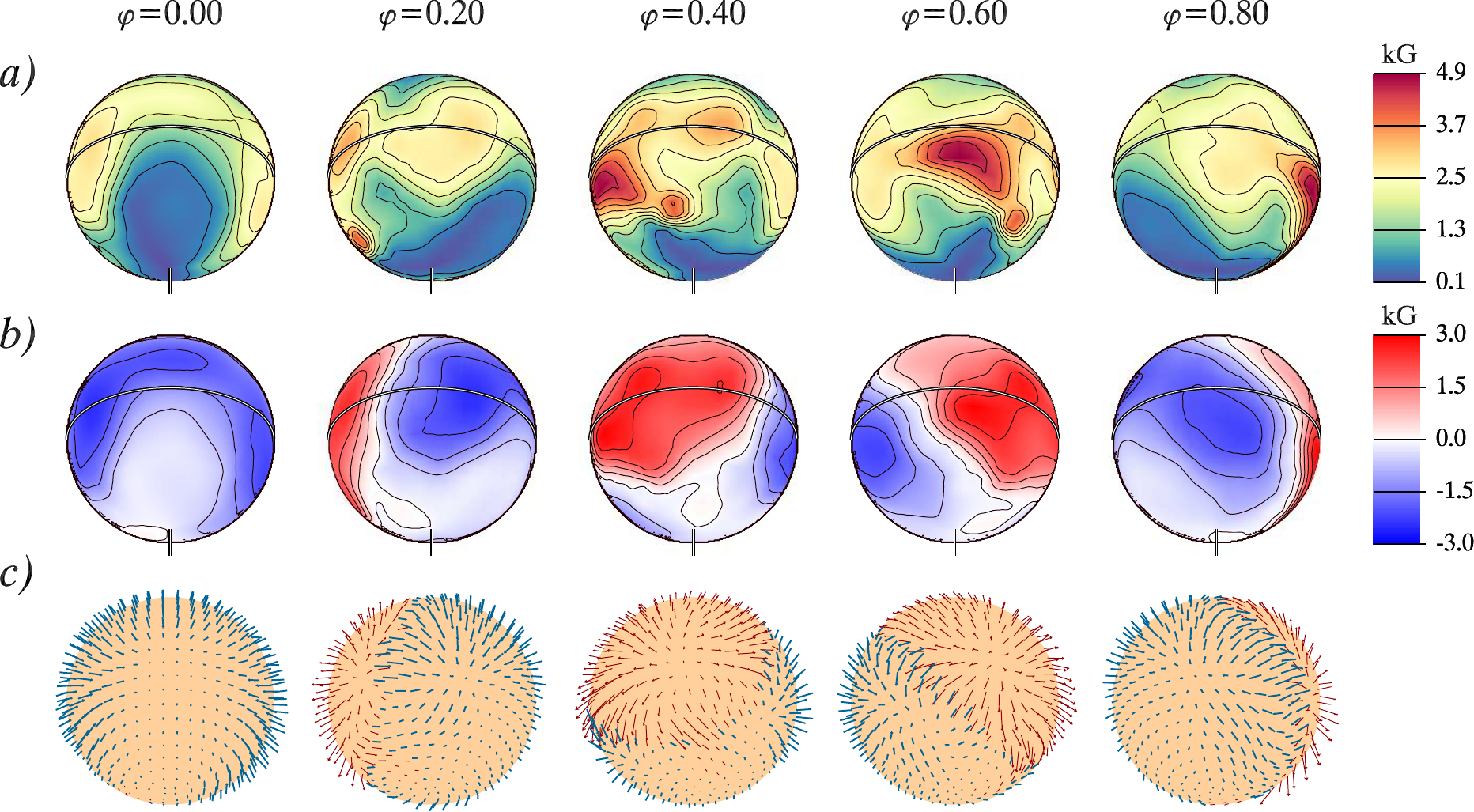}
   \caption{Magnetic map computed using all selected iron lines and chromium lines from the ESPaDOnS/Narval data. The spherical plots show distributions of the field modulus (a), radial field (b) and field orientation (c).}
\label{finalmapfig}
\end{center}
\end{figure*}

\subsection{Iron and Chromium map - The final map}
By combining all the lines found to be suitable for Stokes $IQUV$ mapping, we produced a magnetic map using iron lines (Fe~{\sc ii}\ $\lambda$ 4923,  5018,  4273,  4520 and 4666), combined with chromium lines (Cr~{\sc ii}\ $\lambda$ 4824, 5246 and 5280).   Because these lines all differ in depth and also magnetic sensitivity,  this will require the code to find a balance between all the competing lines and still produce a consistent map.  It is interesting to test if such a mixture of lines can produce a reasonable result that does not alter the general structure of the field seen in the previous maps. At the same time because of the difference in line depths and respective surface structure sensitivities, this should be considered more representative of the magnetic field topology of $\alpha^2$ CVn.

The fit to the spectra is shown in Fig. \ref{CrFe-fit}. Considering the variety of line depths included in this inversion, very good agreement between the model and the observations can be seen at all phases and in all Stokes parameters.  The magnetic map created with the chromium and iron lines (herein referred to as the final map)  is shown compared with the strong iron line magnetic map along with other line maps in Fig. 3 (map bottom row).  There is a general agreement in the overall structure,  but there is also a difference between the meridional component, with the final map showing the main negative field structure concentrated at a higher latitude compared to the strong iron line map.  This difference is reduced by comparing the final map with the combined iron line map (which includes weak and strong lines) as described in section 4.3.  This indicates the inclusion of weak iron lines makes a contribution to the overall meridional field at negative latitudes, again combating the potential smoothing effect of the strong iron line map.  Whilst still being reasonably consistent with the strong field iron map, we consider this final map to be more representative of the true field, with the smoothing effects of the strong iron lines somewhat limited.  This final map is shown in spherical form in Fig. \ref{finalmapfig}, with field modulus (a), radial field (b) and field orientation (c) shown. 

Figs. 1, 3 and 6 show that the inversion of the Stokes profile consistently recovers structures on a variety of scales. We can roughly estimate the formal resolution of the maps provided by rotational broadening by dividing twice \vs\ by the spectral resolution. For a resolving power of 65000 and taking the relatively low \vs\ of $\alpha^2$ CVn, we obtain approximately 15 resolution elements along the equator, corresponding to about $24^{\circ}$. However, at low \vs\ values (such as that of $\alpha^2$ CVn) information from rotational modulation becomes progressively more important than information coming from the Doppler broadening.  Ultimately, there is no standard method to estimate resolution from rotational modulation.  A rough lower limit to the achievable spatial resolution can be calculated using average phase sampling (in this case $360^{\circ}$ / 28 phases = $13^{\circ}$). It is worth noting that our maps don't appear to show any consistent structures smaller than $20^{\circ}$.

\section{Conclusion}
With the results presented in section 4.1 we are able to show that the magnetic maps from K\&W and magnetic maps created using the new observations of Silvester et al. (2012) are indeed consistent when considering the iron lines Fe~{\sc ii}\ $\lambda$ 4923 and 5018, with only small differences likely attributable to the difference in data quality (spectral resolution and signal-to-noise) between the two data sets. This also illustrates that the small scale structure is stable over the timescale between the two epochs of observation. 

In section 4.2 we investigated the potential impact of line selection on the resulting magnetic map by using only weak iron lines in the inversion.  This produced a magnetic map that was broadly consistent with the strong line map and only differed in the meridional field at low latitudes and the azimuthal field at high latitudes. The largest difference in the azimuthal field is on the order of 1 kG and located around a latitude of $25^{\circ}$. The difference in the meridional field is on the order of 1.5 kG and located around a latitude of $-60^{\circ}$. Both differences are located around a longitude of $180^{\circ}$. This difference is believed to be a result of the difference in horizontal structure sensitivities between the weak and strong line sets, with stronger saturated lines altered less by abundance spots. It is important to note that profiles computed from weak lines did not agree well with profiles computed from strong iron lines. In addition the meridional field is the most difficult to constrain by inversion. This suggests the strong iron map represents a smoothed version of the magnetic field map. 

It should be noted that a potential source of the discrepancy seen between the weak and the strong iron magnetic maps is vertical stratification.  Theoretical modelling of abundance stratification by Alecian and Stift (2010) showed that in an Ap star, vertical stratification leads to a change in chemical abundance as a function of optical depth in the atmosphere. In such a framework, strong lines which sample a larger range of optical depths when compared to weak lines, would be probing different parts of this ``abundance vs optical depth'' variation.  This effect is more significant in cooler Ap stars,  however for  hotter stars like $\alpha^2$ CVn,  Stift and Alecian (2012) showed it could potentially lead to variations in abundances on the order of 1 dex over the range of optical depths. 

By combining strong and weak iron lines the resulting ``combined iron line'' map was slightly more consistent with the strong iron line map, than when weak iron lines alone were used in the mapping. The contribution of the weak lines into the combined iron line map still results in a difference in the meridional field at low latitudes. It was possible to reproduce both the weak and strong iron line profiles with this single map.  In section 4.4 we produced a magnetic map based on chromium lines. The results were similar to those found by mapping weak iron lines, with an overall agreement with the strong iron line map but still with differences in the meridional fit at low latitudes.  Interestingly both the weak iron line map and the chromium lines independently give very similar magnetic field maps. This illustrates that very different line sets can give consistent mapping results. 

Finally in section 4.6 we combined all the above line sets into a final map of the magnetic field of $\alpha^2$ CVn. We consider this map to be the best existing representation of the magnetic field of $\alpha^2$ CVn. The overall structure of the final map is in agreement with K\&W: we find a dipole-like structure with complex sub-structure. As was found by K\&W, we find that the magnetic field is strongest at the positive pole (seen clearly at phase 0.6), with an asymmetry compared to the negative pole.  

There is clear agreement between the maps of K\&W and our new maps (produced from a completely new set of Stokes $IQUV$ observations). When this is combined with the fact that we are able to reproduce the same general magnetic field topology from a variety of lines sets, of varying intensities and from two different atomic species, it provides compelling support to the findings of Kochukhov et al. (2012), and suggests the magnetic field structure we reconstruct for  $\alpha^2$ CVn is accurate and not the result of limitations in the inversion technique as has been suggested by Stift et al. (2012). In addition we have also illustrated that the observed profiles cannot be fit with a simple dipolar or dipole + quadrupole geometry and can only be fit with direct magnetic mapping using only local regularization to constrain the map. One other important result from this work is that we have the first confirmation via magnetic Doppler imaging that the global magnetic field of $\alpha^2$ CVn is stable over the period of a decade, which adds further evidence to the current theoretical understanding of the stability of the magnetic fields in Ap stars. 

It should be noted that regardless of the line-set choice, the large-scale structure of all the maps is consistent with the original maps of K\&W and strong iron line maps. Line selection may have an subtle effect on the resulting magnetic field map, with small differences arising as a result of horizontal structure sensitivity of the atomic line used in the inversion. With this in mind, it is clearly of value to map a variety of lines with different formation heights when possible, such as weak and strong iron lines  or strong iron lines and chromium lines, etc.  Even when it is not possible to have such a variety,  provided that the line profiles used are of sufficient quality and show clear linear polarisation amplitudes, the reconstructed field should be reliable.  It could be argued that data quality is a far bigger factor potentially affecting the reconstructed magnetic topology.  

The next stage in the project is to produce a series of abundance maps for $\alpha^2$ CVn, looking for any interplay between the magnetic field and chemical atmospheric structures.  We will then investigate the magnetic field geometry and chemical abundance structures of other Ap stars using data from Silvester et al. (2012) and the MiMeS project.  By increasing the sample of Ap stars studied using MDI, we can probe what influences other stellar parameters (such as mass, temperature, rotation etc) have on the resulting magnetic field geometry.

\section*{Acknowledgments} 
OK is a Royal Swedish Academy of Sciences Research Fellow supported by grants from the Knut and Alice Wallenberg Foundation,  the Swedish Research Council and G\"{o}ran Gustafsson Foundation.
GAW acknowledges support from the Natural Science and Engineering Research Council of Canada in the form of a Discovery Grant.
JS thanks Dr Kristine Spekkens for providing a workstation on which to run INVERS10

\bsp

\begin{landscape}
\vspace*{15 mm}
\begin{figure}
\includegraphics[width=0.62\textwidth]{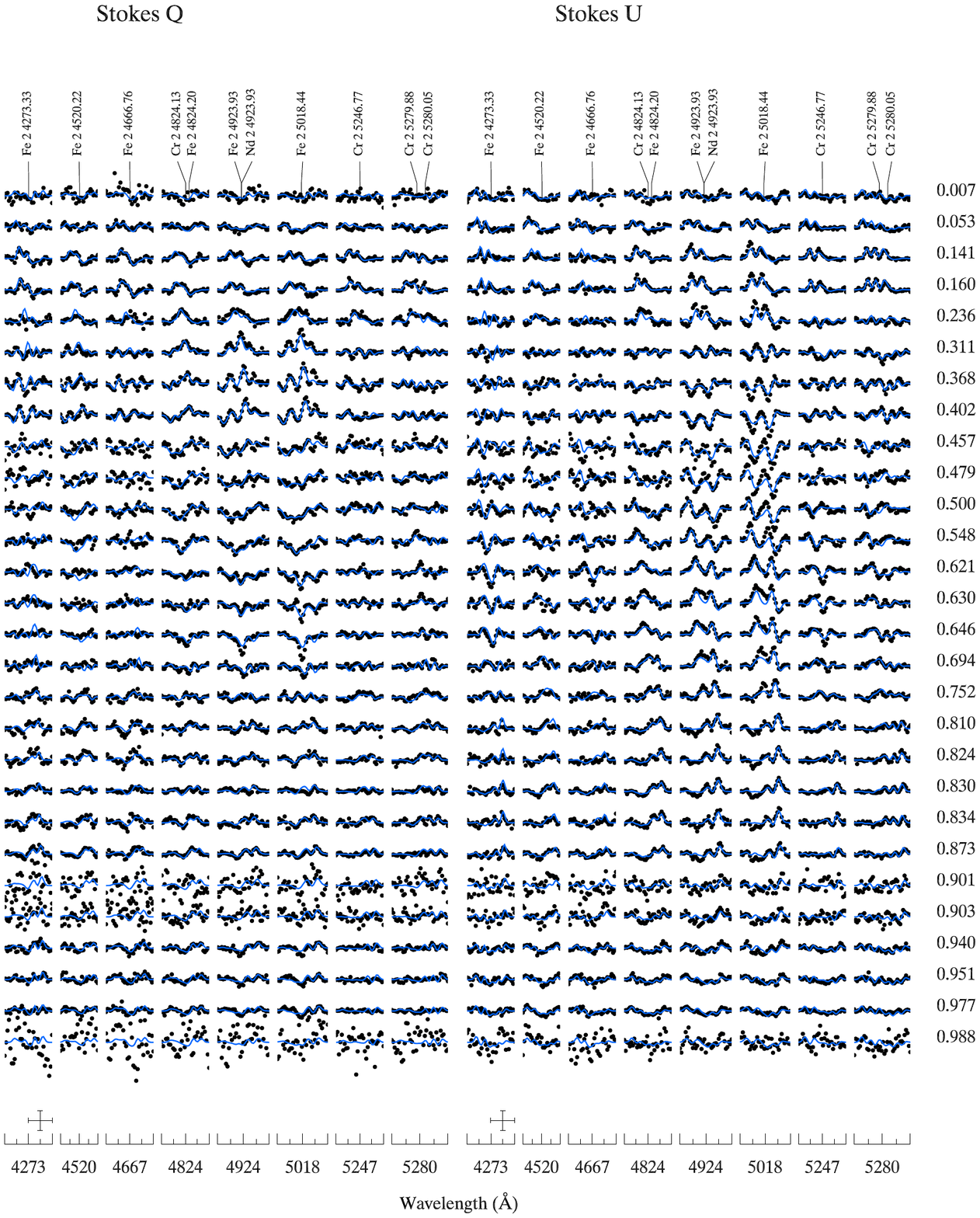}
\includegraphics[width=0.62\textwidth,]{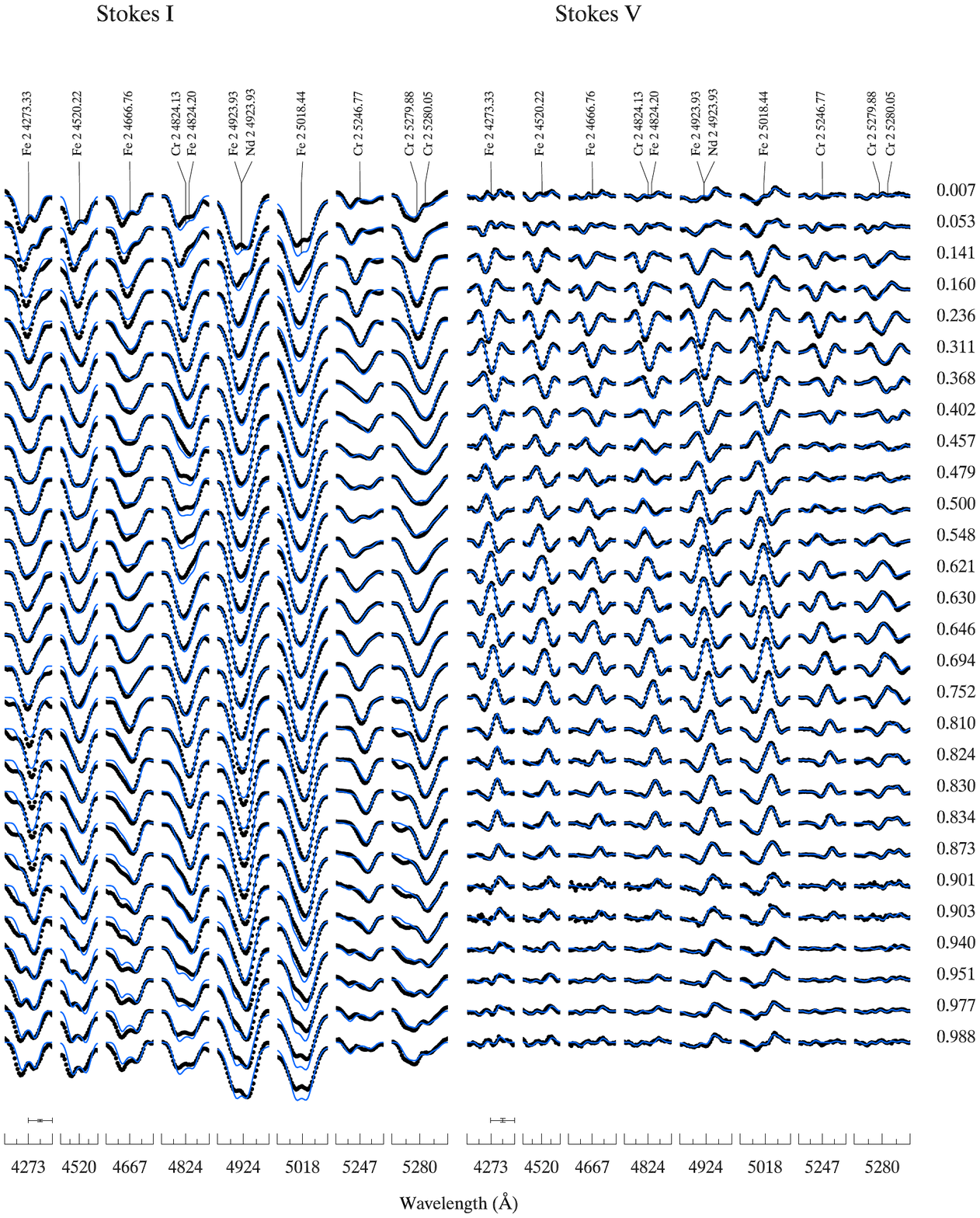}
 \caption{Comparison between observed (dots) and synthetic (solid curve, blue) Stokes $IQUV$ parameter spectra of $\alpha^2$ CVn for the final magnetic map (iron and chromium combined). The solid curves show the best fit to all Cr and Fe lines combined. }
\label{CrFe-fit}
\end{figure}
\end{landscape}



\begin{thebibliography}{1}
\bibitem[{Alecian G., Stift M. J.}, 2010]{Ale} {Alecian G., Stift M. J.}, 2010, A\&A, 516, 53
\bibitem[Babcock H. W and Burd S., 1952]{bab} Babcock H. W and Burd S., 1952, ApJ, 116, 8
\bibitem[Bagnulo S., Landi Degl'Innocenti M., Landolfi M., Mathys, G., 2002]{ban}Bagnulo S., Landi Degl'Innocenti M., Landolfi M., Mathys, G.,2002, A\&A, 394, 1023
\bibitem[Borra E.F., and Landstreet J.D., 1977]{borra1} Borra E.F., and Landstreet J.D., 1977,  ApJ,  212, 141
\bibitem[Borra E.F., and Vaughan A.H., 1978]{borra2} Borra E.F., and Vaughan A.H., 1978, ApJ, 220, 924
\bibitem[Brown S. F., Donati  J.-F., Rees D. E., \& Semel, M., 1991]{brown}Brown S. F., Donati  J.-F., Rees D. E., \& Semel, M., 1991, A\&A, 250, 463
\bibitem[Donati J.-F., et al., 2006]{don} Donati J.-F., et al., 2006, MNRAS, 370, 629
\bibitem[Farnsworth G., 1932]{farn} Farnsworth G., 1932, ApJ, 76, 313
\bibitem[Kochukhov O., Bagnulo S., Wade G. A., Sangalli L., Piskunov N., Landstreet J. D., Petit P., Sigut T. A. A.,  2004]{koch} Kochukhov O., Bagnulo S., Wade G. A., Sangalli L., Piskunov N., Landstreet J. D., Petit P., Sigut T. A. A.,  2004, A\&A, 414, 613
\bibitem[Kochukhov O., Piskunov N., Ilyin I., Ilyina S., Tuominen, I., 2002]{koch2} Kochukhov O., Piskunov N., Ilyin I., Ilyina S., Tuominen, I., 2002, A\&A 389, 420
\bibitem[Kochukhov O. \& Piskunov N., 2002]{koch3} Kochukhov O. \& Piskunov N., 2002, A\&A 288, 868
\bibitem[Kochukhov O., Bagnulo S., 2006]{kb}Kochukhov O., Bagnulo S., 2006, A\&A, 450, 763
\bibitem[Kochukhov O. and Wade, G.A, 2010]{kw}Kochukhov O. and Wade, G.A, 2010, A\&A, 513, A13
 \bibitem[Kochukhov O., Wade G. A., Shulyak D., 2012]{kw2}Kochukhov O., Wade G. A., Shulyak D., 2012, MNRAS, 421, 3004
 \bibitem[Kochukhov O., Mantere M. J., Hackman T., Ilyin I., 2013]{koch4}Kochukhov O., Mantere M. J., Hackman T., Ilyin I., 2013, A\&A, 550, 84
\bibitem[Kupka F., Piskunov N.,  Ryabchikova T. A., Stempels H. C., Weiss W. W., 1999]{kup}[Kupka F., Piskunov N.,  Ryabchikova T. A., Stempels H. C., Weiss W. W., 1999, A\&A, 138, 119
\bibitem[Markov A., 1930]{Mar}Markov A., 1930, ApJ, 72, 301
 \bibitem[Piskunov N., Kochukhov O., 2002]{Pis} Piskunov N., Kochukhov O., 2002, A\&A, 381. 736
 \bibitem[Pyper D.M., 1969]{Pyp}Pyper D.M., 1969, ApJS, 18, 347
 \bibitem[Ros\'{e}n L. ,  Kochukhov O., 2012]{Ros}[Ros\'{e}n L. ,  Kochukhov O., 2012, A\&A, 548, A8
 \bibitem[Silvester J., Wade G.A., Kochukhov O., S. Bagnulo S., Folsom C.P., Hanes D., 2012]{sil} Silvester J., Wade G.A., Kochukhov O., S. Bagnulo S., Folsom C.P., Hanes D., 2012, MNRAS,  426, 1003
 \bibitem[Stift M. J., Leone F., Cowley C. R., 2012]{sti} Stift M. J., Leone F., Cowley C. R., 2012, MNRAS. 419, 2912
  \bibitem[Stift M. J., Alecian G., 2012]{sti2}Stift M. J., Alecian G., 2012, MNRAS, 425, 2715
 \bibitem[Wade G.A., Donati J.-F., Landstreet J.D., Shorlin S.L.S., 2000]{wad}Wade G.A., Donati J.-F., Landstreet J.D., Shorlin S.L.S., 2000, MNRAS, 313, 823
\end{thebibliography}
\end{document}